\documentclass[aps,pra,twocolumn,groupedaddress,showpacs,superscriptaddress]{revtex4}
\usepackage{graphicx}

\begin{document}

\title{Quantum phases of mixtures of atoms and molecules on optical lattices}

\author{V.G.~Rousseau}
\author{P.J.H.~Denteneer}
\affiliation{Instituut Lorentz, Universiteit Leiden, Postbus 9506, 2300 RA Leiden, The Netherlands}

\begin{abstract}
We investigate the phase diagram of a two-species Bose-Hubbard model including a conversion term,
by which two particles from the first species can be converted into one particle of the second species, and vice-versa. The model
can be related to ultracold atom experiments in which a Feshbach resonance produces long-lived
bound states viewed as diatomic molecules. The model is solved exactly by means of Quantum Monte Carlo simulations.
We show that an "inversion of population" occurs, depending on the parameters, where the second species becomes more numerous than the first species. The model also exhibits an exotic incompressible "Super-Mott" phase where the particles from both species can flow with signs of superfluidity,
but without global supercurrent. We present two phase diagrams, one in the (chemical potential, conversion)-plane, the other in the (chemical potential, detuning)-plane.
\end{abstract}

% 05.30.Jp = Boson systems
% 03.75.Hh = Static properties of condensates; thermodynamical,
%            statistical and structural properties
% 67.40.Kh = Boson degeneracy and superfluidity of He4: Thermodynamic properties
% 75.10.Nr  Spin-glass and other random models
% 71.10.Fd  Lattice fermion models (Hubbard model, etc.)
% 71.30.+h  Metal-insulator transitions and other electronic transitions
% 02.70.Uu  Applications of Monte Carlo methods
\maketitle

\section{Introduction}
In the past years the Bose-Hubbard model \cite{Fisher89} has been extensively investigated and a lot of interest has been generated thanks
to ultracold atom experiments on optical lattices \cite{Jaksch98}, which provide an ideal realization of the model. Recently, much
theoretical and experimental work has been performed on mixtures with several species of particles. For instance, Bose-Fermi mixtures
on lattices have been studied \cite{Ott04,Gunter06,Ospelkaus06,Pollet06,Sengupta07,Hebert06}. Another mixture that is likely of interest involves atoms and molecules, in which conversion between the two species is possible.
Such conversion processes can describe, for instance, long-lived bound states of atoms (diatomic molecules)
occuring in ultracold atom experiments where a Feshbach resonance is used to tune the scattering length of the atoms \cite{Timmermans,Dickerscheid}. In those experiments, the hyperfine interaction
between two spin polarized atoms can flip the spin of one of the atoms, reducing sensitively their scattering length.
The two atoms are virtualy bound into a "molecular" state until the hyperfine interaction flips again the spin of one of the atoms.

\section{The model}
With the motivation above,
we propose to study a two-boson species model with an additional conversion term allowing two particles from the first species to turn
into one particle of the second species, and vice-versa.
We denote the first species as "atoms",
and the second species as (diatomic) "molecules". Atoms and molecules can hop onto neighboring sites, interact, and conversion between two atoms and a
molecule can occur. Several atoms can reside on the same site, their interaction being described by an on-site repulsion potential. A second on-site repulsion potential describes the interactions between molecules and atoms being on the same site. This leads us to consider the following Hamiltonian
\begin{eqnarray}
  \label{Hamiltonian} \hat H=\hat T+\hat P+\hat C,
\end{eqnarray}
with
\begin{eqnarray}
  \label{Kinetic}    && \!\!\!\!\!\!\hat T\!=\!-t_a\!\!\sum_{\big\langle i,j\big\rangle}\!\!\big(a_i^\dagger a_j^{\phantom\dagger}+h.c.\big)-t_m\!\!\sum_{\big\langle i,j\big\rangle}\!\!\big(m_i^\dagger m_j^{\phantom\dagger}+h.c.\big) \\
  \label{Potential}  && \!\!\!\!\!\!\hat P\!=\!U_{aa}\!\sum_i\! \hat n_i^a\big(\hat n_i^a\!\!-\!\!1\big)\!+\!U_{am}\!\sum_i\! \hat n_i^a \hat n_i^m\!+\!D\!\sum_i\! \hat n_i^m \\
  \label{Conversion} && \!\!\!\!\!\!\hat C\!=\!g\sum_i\big(m_i^\dagger a_i^{\phantom\dagger} a_i^{\phantom\dagger}+a_i^\dagger a_i^\dagger m_i^{\phantom\dagger}\big).
\end{eqnarray}

The $\hat T$, $\hat P$, and $\hat C$ operators correspond respectively to the kinetic, potential, and conversion energies. The $a_i^\dagger$ and $a_i^{\phantom\dagger}$ operators ($m_i^\dagger$ and $m_i^{\phantom\dagger}$) are the creation and annihilation operators of atoms (molecules)
on site $i$, and $\hat n_i^a=a_i^\dagger a_i^{\phantom\dagger}$ ($\hat n_i^m=m_i^\dagger m_i^{\phantom\dagger})$ counts the number of atoms (molecules) on site $i$. Those
operators satisfy the usual bosonic commutation rules $\big[a_i^{\phantom\dagger},a_j^\dagger\big]=\delta_{ij}$ and $\big[m_i^{\phantom\dagger},m_j^\dagger\big]=\delta_{ij}$.
In order to simplify the model and reduce the space of parameters, we impose a hard-core constraint on molecules. This is done by adding the condition \mbox{$m_i^{\phantom\dagger} m_i^{\phantom\dagger}=m_i^\dagger m_i^\dagger=0$}. For a minimal model, we set a maximum of two atoms per site by imposing \mbox{$a_i^{\phantom\dagger} a_i^{\phantom\dagger} a_i^{\phantom\dagger}=a_i^\dagger a_i^\dagger a_i^\dagger=0$}. The sums $\big\langle i,j\big\rangle$ run over pairs of nearest-neighboring sites $i$ and $j$. We restrict our study to one dimension and we choose the atomic
hopping parameter $t_a=1$ in order to set the energy scale, while we choose the molecular hopping parameter $t_m=1/2$, motivated by the continuous-space behavior of the hopping as a function of the mass ($t\propto\hbar^2/2m$), a molecule being twice heavier than an atom. Smaller values of $t_m$ (as mapping of experimental systems to Bose-Hubbard models would suggest \cite{Timmermans,Dickerscheid}) are not expected to lead to qualitatively different behavior. The parameter $U_{aa}$ controls the interaction strength between atoms, and $U_{am}$
controls the interaction between atoms and molecules. The conversion between atoms and molecules is controlled by the positive parameter $g$. This parameter can be related to the
"hyperfine interaction" parameter in the Feshbach resonance picture \cite{Timmermans,Dickerscheid}. Finally, the parameter $D$ acts as a chemical potential for
molecules, and allows to tune the energy difference between atomic and molecular states. This parameter can be related to the "detuning"
in the Feshbach resonance example. In the remainder of the paper we will not expand on the connection to the Feshbach resonance problem, nor attempt to reproduce Feshbach resonance physics. We concentrate on taking the model given in (\ref{Hamiltonian})--(\ref{Conversion}) at face value and determining its phase diagram. A similar model for the one-dimensional continuum has been analysed in Ref \cite{Gurarie}, and
for optical lattices in the mean field approximation \cite{Dupuis}.

It is important to note that the Hamitonian (\ref{Hamiltonian}) does not conserve the number of atoms
$N_a=\sum_i a_i^\dagger a_i^{\phantom\dagger}$, nor the number of molecules $N_m=\sum_i m_i^\dagger m_i^{\phantom\dagger}$, because of the
conversion term (\ref{Conversion}). However we consider that a molecule is made of two particles, so the total number of particles $N$ in the system is conserved:
\begin{equation}
  \label{TotalNumber} N=N_a+2N_m.
\end{equation}

\section{Quantum Monte Carlo simulations: The World Line algorithm}
\begin{figure}[h]
  \centerline{\includegraphics[width=0.45\textwidth]{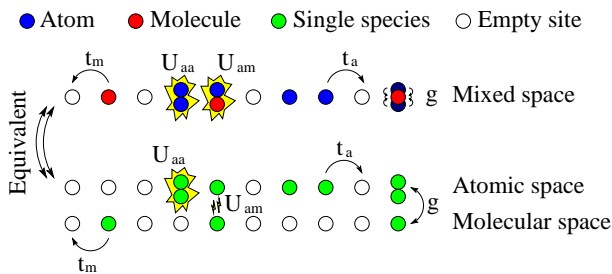}}
  \caption
    {
      (Color online) In order to make the model suitable for simulations, a mapping is performed between the model of atoms
      and molecules living on a 1D lattice, and a model of single species where the particles reside on a ladder.
    }
  \label{Mapping}
\end{figure}

In order to make the model suitable for simulations, we perform a mapping of the Hamiltonian describing two species of bosons on a
1D lattice (\ref{Hamiltonian}) onto a Hamiltonian describing single species of bosons evolving on a ladder (Fig.~\ref{Mapping}).
In the 1D space, the two species live together. They can hop onto neighboring sites, and the interaction between the two species is
described by an on-site potential $U_{am}$. The conversion between the two species occurs on a single site. In the ladder space, the
atoms (molecules) live on the top (bottom) side of the ladder. The interaction between the two species is described by a
potential $U_{am}$ acting between vertical neighboring sites. Two atoms living on the same atomic site can be destroyed at the same time, with
the creation of a molecule on the corresponding molecular site (and vice-versa).

Quantum Monte Carlo simulations are performed for the ladder model by making use of the World Line algorithm \cite{Batrouni1990,Batrouni1992}. It is essential to
emphasize that this algorithm works in the canonical ensemble, meaning here that the total number of particles $N=N_a+2N_m$ is conserved.
Indeed, simulations using a grand canonical algorithm (Stochastic Series Expansion) \cite{Sandvik} turned out to be difficult to handle, because it is numerically very hard to control
the number of particles of each species using two chemical potentials, the number of particles of each species depending on both chemical potentials.

Defining the continuous product of evolution operators in imaginary time,
\begin{eqnarray}
  \prod_{0\to\beta}^{d\tau}e^{-d\tau\hat H}\hat=\lim_{M\to\infty}\prod_{k=1}^M e^{-\frac{\beta}{M}\hat H}=e^{-\beta\hat H},
\end{eqnarray}
one starts by writing the partition function as the trace of the evolution operator $e^{-\beta\hat H}$
\begin{eqnarray}
  \mathcal Z=\sum_{\psi}\left\langle\psi\right|\prod_{0\to\beta}^{d\tau} e^{-d\tau\hat H}\left|\psi\right\rangle
\end{eqnarray}
using the occupation number representation for the states $\left|\psi\right\rangle$. Then we use the so-called "checkerboard decomposition"
for the Hamiltonian, $\hat H=\hat H_e+\hat H_o$, with
\begin{eqnarray}
  \hat H_e=\sum_{i\,even}\hat H_i \hspace{2cm} \hat H_o=\sum_{i\,odd}\hat H_i,
\end{eqnarray}
where $\hat H_i=\hat T_i+\frac12\hat P_i+\frac12\hat C_i$ and $\hat T_i,\hat P_i,\hat C_i$ are defined by
\begin{eqnarray}
            \hat T_i & = & -t_a\big(a_i^\dagger a_{i+1}^{\phantom\dagger}+h.c.\big)-t_m\big(m_i^\dagger m_{i+1}^{\phantom\dagger}+h.c.\big) \\
            \hat P_i & = & U_{aa}\big[\hat n_i^a\big(\hat n_i^a-1\big)+\hat n_{i+1}^a\big(\hat n_{i+1}^a-1\big)\big] \\
  \nonumber          & + & U_{am}\big(\hat n_i^a \hat n_i^m+\hat n_{i+1}^a \hat n_{i+1}^m\big) \\
  \nonumber          & + & D\big(\hat n_i^m+\hat n_{i+1}^m\big) \\
\label{Sign} \hat C_i & = & -g\big(m_i^\dagger a_i^{\phantom\dagger} a_i^{\phantom\dagger}+a_i^\dagger a_i^\dagger m_i^{\phantom\dagger}\big) \\
  \nonumber          & - & g\big(m_{i+1}^\dagger a_{i+1}^{\phantom\dagger} a_{i+1}^{\phantom\dagger}+a_{i+1}^\dagger a_{i+1}^\dagger m_{i+1}^{\phantom\dagger}\big).
\end{eqnarray}

We attract here the attention of the reader to Eq.\ref{Sign}, in which we have added a minus sign to the conversion term. The energy of the model is independent
of the sign in (\ref{Sign}), so (\ref{Sign}) and (\ref{Conversion}) are equivalent. This can be seen by realizing that flipping the sign of the conversion term just results in a redefinition
of the phase of the molecular creation and annihilation operators, $m_i^{\dagger\prime}=-m_i^\dagger$
and $m_i^\prime=-m_i$. We work with a minus sign in (\ref{Sign}) in order to ensure that all matrix elements $\big\langle\phi\big|e^{-\tau\hat\mathcal H}\big|\psi\big\rangle$ are positive.
Those positive matrix elements normalized by $\mathcal Z$ define the probability of transition from the state $\big|\psi\big\rangle$ to the state $\big|\phi\big\rangle$, which is required for a Monte Carlo sampling.

It is important to note that $\hat H_e$ and $\hat H_o$ are written each as a sum of operators $\hat H_i$ that commute (but $\hat H_e$ and
$\hat H_o$ do not commute). Using the Trotter-Suzuki formula at second order,
\begin{eqnarray}
  e^{-d\tau\big(\hat H_e+\hat H_o\big)}=e^{-\frac12 d\tau\hat H_o}e^{-d\tau\hat H_e}e^{-\frac12 d\tau\hat H_o}+\mathcal O(d\tau^3),
\end{eqnarray}
and using properties of the trace we get
\begin{eqnarray}
  \mathcal Z=\sum_\psi\left\langle\psi\right|\prod_{0\to\beta}^{d\tau}e^{-d\tau\hat H_e}e^{-d\tau\hat H_o}\left|\psi\right\rangle.
\end{eqnarray}

The error due to the Trotter-Suzuki decomposition vanishes because of the continuous product making $d\tau$ going to zero (in the
case of a discrete product the Trotter error becomes $\mathcal O\big(d\tau^2\big)$ instead of $\mathcal O\big(d\tau^3\big)$, due to the accumulation of errors in the
product). Introducing complete sets of states $I=\sum_{\psi(\tau)}\left|\psi(\tau)\right\rangle\left\langle\psi(\tau)\right|$ between
each pair of exponentials leads to
\begin{eqnarray}
  \label{Partition} \!\!\!\mathcal Z=\!\!\!\sum_{[\psi(\tau)]_0^\beta} \prod_{0\to\beta}^{d\tau} \left\langle\psi(\tau\!\!+\!\!d\tau)\right|e^{-d\tau\hat H_e}\left|\psi(\tau\!\!+\!\!d\tau/2)\right\rangle \\
  \nonumber         \!\!\!                                         \times              \left\langle\psi(\tau\!\!+\!\!d\tau/2)\right|e^{-d\tau\hat H_o}\left|\psi(\tau)\right\rangle,
\end{eqnarray}
where the sum runs over all sets of states $\psi(\tau)$ for all values of $\tau$ in $[0,\beta]$. Finally, each operator $e^{-d\tau\hat H_e}$
and $e^{-d\tau\hat H_o}$ is a product of independent four-site operators $e^{-d\tau\hat H_i}$ (2 sites $i$ and $i+1$ in the atomic
space and 2 sites in the molecular space). With the hard-core constraint on molecules and a maximum of two atoms per site, the size of the
Hilbert space of the four-site problem is 36. Thus each matrix element in (\ref{Partition}) can be computed by evaluating numerically
$36\times 36$ matrices. As a result, the quantum problem has been mapped onto a classical problem with an extra imaginary time dimension, and
the algorithm consists in generating configurations of states $\psi(\tau)$ using standard classical Monte Carlo techniques. For more details, see references \cite{Batrouni1992,Rousseau2005}.

\section{Quantities of interest}
In addition to the atomic and molecular densities,
\begin{eqnarray}
  \rho_a=N_a/L \hspace{1cm} \rho_m=N_m/L,
\end{eqnarray}
we also define the total density
\begin{eqnarray}
\rho_{tot}=\frac{N_a+2N_m}{L},
\end{eqnarray}
by analogy with (\ref{TotalNumber}), where $L$ is the number of sites in the lattice.

In order to identify insulating phases, it is useful to look at the behavior of the total density $\rho_{tot}$ as a function of the chemical
potential $\mu(N)$. It is common to define the chemical potential in the canonical ensemble at zero temperature by the energy cost to add one particle to the
system, \mbox{$\mu(N)=E(N+1)-E(N)$}. However, for our present model, it is better to define it by the energy cost to add successively 2 particles to the system
divided by 2,
\begin{eqnarray}
  \mu(N)=\frac{E(N+2)-E(N)}{2}.
\end{eqnarray}
Indeed, this allows to keep an even total number of particles, preventing an extra single particle to
be out of the atoms/molecules conversion process.

\begin{figure}[h]
  \centerline{\includegraphics[width=0.45\textwidth]{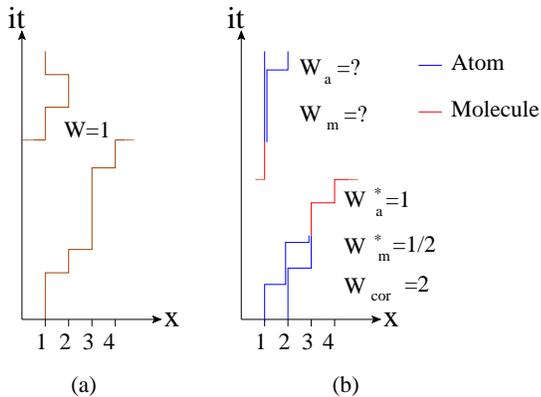}}
  \caption
    {
      (Color online) Example of world lines for a four-site lattice with periodic boundary conditions. (a) For a system without conversion between the different species, the world lines are continuous and the winding number is well-defined. (b) The conversion between atoms and
      molecules leads to discontinuities in the world lines, and no true winding can be defined for each of the species. However it is
      well-defined for the mixture atoms/molecules, because the composite world lines are continuous (see text for details).
    }
  \label{MixtureWinding}
\end{figure}

Another quantity of interest for the characterization of a phase is the superfluid density. An easy way to access this quantity is to make
use of Pollock and Ceperley's formula \cite{PollockCeperley1987} that relates the superfluid density to the fluctuations of the winding number $W$,
$\rho_s=L\big\langle W^2\big\rangle/2t\beta$, where $t$ is the hopping of the considered species, $\beta$ the inverse temperature, and $L$ the number of lattice sites. Usually, this
winding number $W$ is perfectly well-defined for systems with $n$ species of particles. For a given configuration, it is defined by the number of times that the world lines cross the boundaries of the system from the left to the right, minus
the number of times they cross the boundaries from the right to the left (Fig.~\ref{MixtureWinding}a). But in our case, the atomic and
molecular windings, $W_a$ and $W_m$, are ill-defined because the world lines associated to each of the species may be discontinous if
conversions between atoms and molecules occur (Fig.~\ref{MixtureWinding}b). It is then no longer possible to determine whether a particle
is flowing to the right or to the left as a function of imaginary time. However we can define atomic and molecular pseudo-windings, $W_a^\star$ and $W_m^\star$,
by the number of right jumps minus the number of left jumps, normalized by the number of sites $L$. Non-zero values of such pseudo-windings
are signatures of superfluidity of the particles.  When no conversion between atoms and molecules occurs, the definition of pseudo-winding coincides with that of true winding. In addition,
the \textit{correlated} winding is well-defined for the mixture of particles,
\begin{eqnarray}
  \label{CorrelatedWinding} W_{cor}=W_a^\star+2W_m^\star,
\end{eqnarray}
because the composite atomic and molecular world lines are continuous (if one considers that a molecular
world line represents two atomic world lines). This correlated winding is relevant for the
superfluid density of the mixture because it corresponds to the winding of particles, without looking at their individual nature (atom or
molecule). It is also interesting to consider the \textit{anti-correlated} winding,
\begin{eqnarray}
  \label{AnticorrelatedWinding} W_{ant}=W_a^\star-2W_m^\star
\end{eqnarray}
which allows to determine if atoms and molecules are flowing in opposite directions or not. The definitions of correlated winding (\ref{CorrelatedWinding})
and anti-correlated winding (\ref{AnticorrelatedWinding}) are similar to those used in Bose-Fermi mixtures \cite{Pollet06,Hebert06}.

\section{Numerical results}
\subsection{The one-site problem}
\begin{figure}[h]
  \centerline{\includegraphics[width=0.45\textwidth]{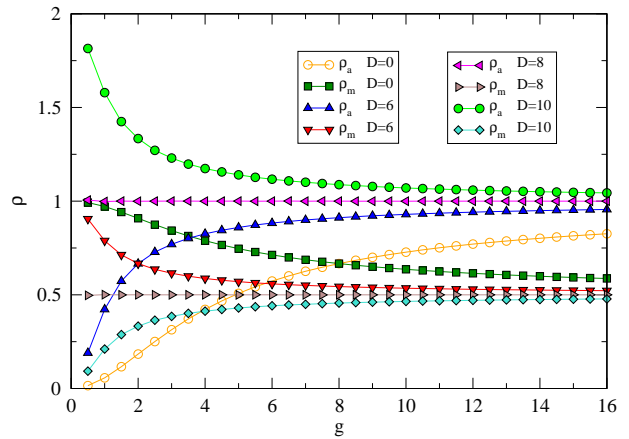}}
  \caption
    {
      (Color online) The one-site problem with 2 particles. The densities of atoms and molecules are plotted as functions of the conversion parameter $g$
      for differents values of the detuning $D$, for $U_{aa}=4$.
    }
  \label{RhoVsg1}
\end{figure}

It is useful to start the investigation of the model by considering first the one-site problem with a total number of particles $N=2$ ($\rho_{tot}=2$).
Figure \ref{RhoVsg1} shows the atomic and molecular densities as functions of the conversion parameter $g$ and different values of
the detuning $D$ for $U_{aa}=4$ (the value of $U_{am}$ does not play any role since there is only 2 atoms or 1 molecule). For $D=0$ and
small $g$, the 2 particles are mainly bound in the molecular state, because the creation of the molecule has a vanishing energy cost while
having 2 atoms costs $2U_{aa}=8$. As $g$ increases, it becomes energetically favorable to make conversions atoms/molecule, so the atomic
density starts to grow, reducing the molecular density. When $g$ is large, the system maximizes the conversion process. Thus the system is
in the molecular state with 1 molecule half of the time, and in the atomic state with 2 atoms the rest of the time. As a result, the atomic
and molecular densities converge to \mbox{$\rho_a=\rho_{tot}/2=1$} and \mbox{$\rho_m=\rho_{tot}/4=1/2$}. For $D=6$ the same behavior holds, but the molecular density decreases
faster to the large $g$ limit because the energy associated to the molecular state is higher and closer to that of the atomic state. For
$D=10$ we have the inverse behavior, the molecular density increases with $g$ and the atomic density decreases, because it is now cheaper
energetically to have 2 atoms rather than 1 molecule. The transition point between those two cases is $D=8=2U_{aa}$ for which the atomic state has exactly the same
energy as the molecular state. Those states have the same probability, and varying $g$ just changes the rate of conversion between them.
Thus the expectation values of the atomic and molecular densities do not depend on the value of $g$, and remain equal to the values
that optimize the conversion process: $\rho_a=1$ and $\rho_m=1/2$.

\subsection{The lattice problem}
\begin{figure}[h]
  \centerline{\includegraphics[width=0.45\textwidth]{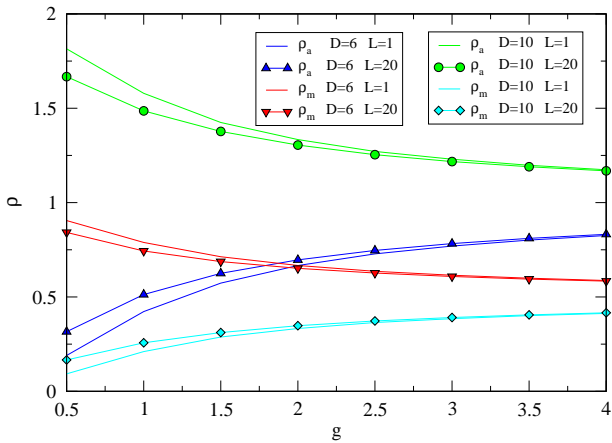}}
  \caption
    {
      (Color online) The atomic and molecular densities as functions of the conversion parameter $g$ and different values of the detuning $D$,
      for $U_{aa}=4$, $U_{am}=12$, and total density of particles $\rho_{tot}=2$.
    }
  \label{RhoVsg2}
\end{figure}

We now turn to the full problem with $L$ lattice sites. We have performed simulations for $L=20,40,80,160$ and determined by extrapolation
to $L=\infty$ that finite size effects associated to the choice of working with \mbox{$L=20$} lead to errors smaller than our statistical error bars,
these latter being smaller than the size of the symbols displayed in the figures of this paper (unless otherwise stated). In the same manner, we have
determined that using $\beta=L$ allows to get the physics relevant to the ground state ($\beta=\infty$), for the measured quantities. As for the one-site problem,
we start by looking at the atomic and molecular densities as functions of $g$, for different values of the detuning $D$, with $U_{aa}=4$,
$U_{am}=12$, and $\rho_{tot}=2$ (Fig. \ref{RhoVsg2}). We can see that going from $L=1$ to $L=20$ (equivalent to turning on the hopping parameters
$t_a$ and $t_m$) just leads to small differences at small $g$. For large $g$ the hopping can be neglected, and results for $L=20$ converge
to those for $L=1$. Nevertheless it is crucial to keep working with the full lattice problem instead of the one-site problem, since this is required to
access global quantities such as the superfluid density. It is also the only way to get results for nearly-continuous values of $\rho_{tot}$.

\begin{figure}[h]
  \centerline{\includegraphics[width=0.45\textwidth]{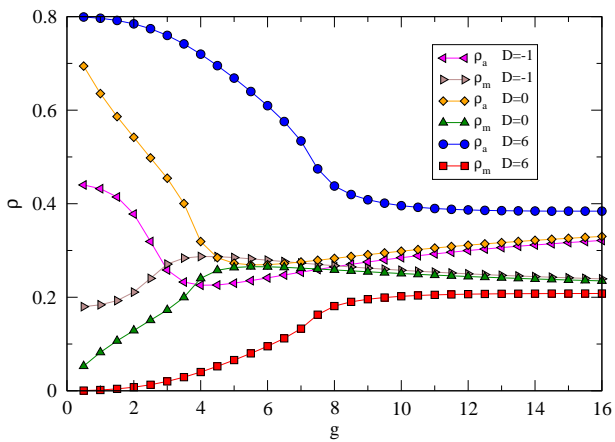}}
  \caption
    {
      (Color online) The atomic and molecular densities as functions of the conversion parameter $g$ at different values of the detuning $D$.
      $U_{aa}=4$, $U_{am}=12$, and total density of particles $\rho_{tot}=4/5$.
    }
  \label{RhoVsg3}
\end{figure}

A completely different behavior occurs when considering a non-commensurate density, for instance $\rho_{tot}=4/5$ (Fig. \ref{RhoVsg3}).
We consider here the case $D<2U_{aa}$ for simplicity. For this density, atoms can be placed on the lattice without increasing the interaction
energy. The same holds for the molecules if $D\leq 0$. But for small $g$ and $D\geq0$ it is energetically more favorable to have atoms only, because 2 atoms have
kinetic energy 4 times more negative than 1 molecule ($2(-t_a)=4(-t_m)$). As a result the molecular density is vanishing for $g=0$ and $D\geq 0$ and grows when
turning on $g$, until reaching the optimal density for large $g$, \mbox{$\rho_m=\rho_{tot}/4=1/5$}, in contrast to Fig. \ref{RhoVsg2}, where for $D\leq 6$ the density $\rho_m$ decreases with increasing $g$. The atomic density follows the inverse
behavior, starts for \mbox{$\rho_a=\rho_{tot}$} and converges to the optimal value, \mbox{$\rho_a=\rho_{tot}/2=2/5$}.

\begin{figure}[h]
  \centerline{\includegraphics[width=0.45\textwidth]{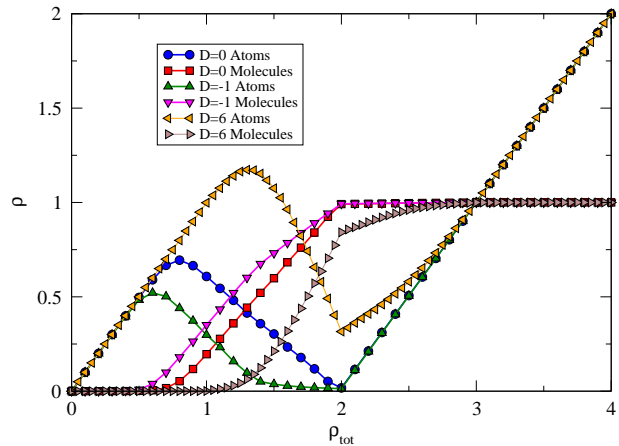}}
  \caption
    {
      (Color online) The densities of atoms and molecules as functions of the total density $\rho_{tot}$ at different values of the
      detuning $D$. $U_{aa}=4$, $U_{am}=12$, and $g=0.5$.
    }
  \label{RhoVsRho}
\end{figure}

Having analyzed the system for two specific values of the total density, it is now interesting to perfom a scan of all values of $\rho_{tot}$. Figure \ref{RhoVsRho} shows the atomic and molecular densities as functions of the total density $\rho_{tot}$. At low filling, the
particles are dilute and the on-site repulsion between atoms prevent double occupancies, so no binding between atoms can occur and the
number of molecules remains zero for all values of $D$ considered. Thus the atomic density increases linearly with the total density. As the filling increases, double
occupancies occur leading to the creation of molecules, and decreasing the atomic density. Increasing the filling further leads to an
"inversion of population" where the number of molecules is greater than the number of atoms. This inversion of population is optimal at
$\rho_{tot}=2$ for the chosen parameters because double atomic occupancies have an energy cost of $2U_{aa}=8$, whereas the creation of a molecule
has an energy cost of $D$. Adding more particles to the system produces a saturation of molecules, and extra atoms just see a constant
potential.

\begin{figure}[ht]
  \centerline{\includegraphics[width=0.45\textwidth]{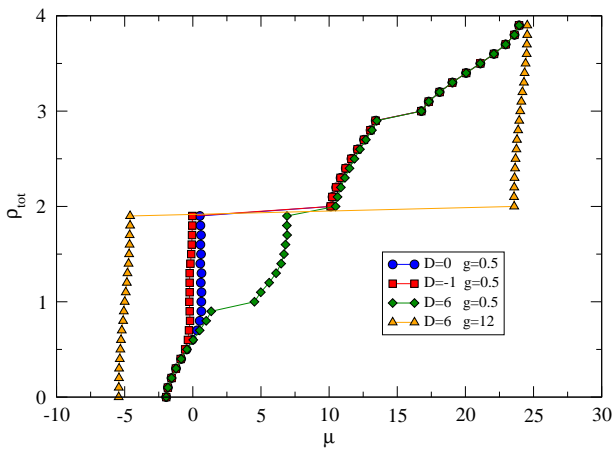}}
  \caption
    {
      (Color online) The total density as a function of the chemical potential, and different values of
      $D$ and $g$, for $U_{aa}=4$, and $U_{am}=12$. The slope of these curves is proportional to the isothermal compressibility,
      and horizontal plateaus indicate phases that are incompressible but not necessarily insulating (see text).
    }
  \label{RhoVsMu}
\end{figure}

In order to identify incompressible phases, it is useful to look at behavior of $\rho_{tot}(\mu)$ for different values of $D$ and $g$ (Fig.  \ref{RhoVsMu}). Let us recall that the slope of this curve, $\partial\rho/\partial\mu$, is
proportional to the isothermal compressibility $\kappa_T$. Thus each horizontal plateau indicates an incompressible Mott phase. This does not
imply that this phase is insulating, as will be shown below. For $D=-1$ or $D=0$ and small conversion $g=0.5$ one can identify two incompressible
phases by the presence of Mott plateaus at $\rho_{tot}=2$ and $\rho_{tot}=3$. For those parameters the usual Mott plateau
occuring in pure bosonic systems at $\rho_{tot}=1$ is absent. This is because extra particles can be added beyond $\rho_{tot}=1$ without
the need of creating double occupancies, by converting atoms into molecules. For $\rho_{tot}=2$, the phase is incompressible
because any site is occupied by a molecule. Thus adding an extra atom requires the formation of an atom/molecule pair, which
has an energy cost of $U_{am}$. For $\rho_{tot}=3$, each site is occupied with an atom/molecule pair, and adding extra atoms leads
to double occupancies with energy costs of $U_{aa}$. Thus the phase is also incompressible. For $D=6$ and $g=0.5$, we recover a Mott plateau at $\rho_{tot}=1$
because creating a molecule has an energy cost of $D$ that cannot be overcome by the associated negative kinetic and conversion energies. For
large $g$ however, the Mott plateaus at $\rho_{tot}=1$ and $\rho_{tot}=3$ disappear. Indeed, in this regime the conversions
between atoms and molecules occur and overcome the energy cost of having two atoms on a single site, as well as the energy cost of creating a
molecule. Thus extra atoms at $\rho_{tot}=1$ and $\rho_{tot}=3$ can go either into a molecule or doubly occupied sites, without changing the
energy by a value greater than the finite-size-lattice gap, which vanishes in the thermodynamic limit. However $\rho_{tot}=2$ is still incompressible because any site is occupied either by two atoms or by a
molecule. Thus an extra atom can go only on a site occupied by a molecule, leading to an energy cost of $U_{am}$. Moreover a conversion process
can no longer take place on this site, and the system has to pay the price of having a molecule all the time with the associated chemical
potential $D$. This explains the large width of the corresponding Mott plateau: approximately $D+g+U_{am}$.

\begin{figure}[ht]
  \centerline{\includegraphics[width=0.45\textwidth]{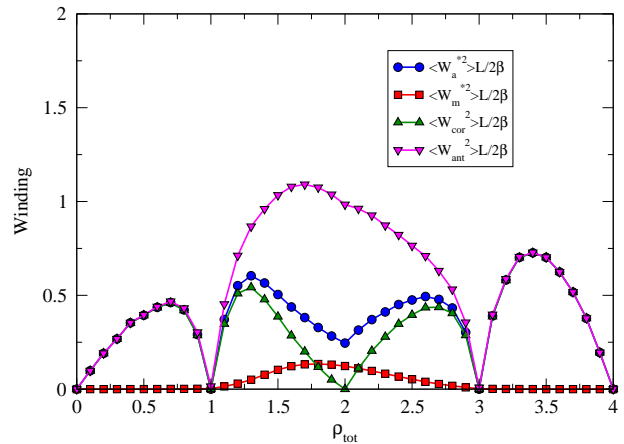}}
  \caption
    {
      (Color online) The winding as a function of the filling for $U_{aa}=4$, $U_{am}=12$, $g=0.5$, and $D=6$. Error bars of the order of the symbol sizes.
    }
  \label{WindingVsRho}
\end{figure}

We now study the potential superfluidity of the mixture by analysing the fluctuations of the atomic and molecular pseudo windings $\big\langle W_a^{\star2}\big\rangle$ and
$\big\langle W_m^{\star2}\big\rangle$, and the correlated and anti-correlated windings $\big\langle W_{cor}^2\big\rangle$ and
$\big\langle W_{ant}^2\big\rangle$ (Fig. \ref{WindingVsRho}), defined in section IV. To discuss the results, it is useful to consider the corresponding curve in Fig. \ref{RhoVsMu} ($D=6$, $g=0.5$; green curve). For $\rho_{tot}=1$ and
$\rho_{tot}=3$ all windings and pseudo windings vanish, showing that the system is frozen for those densities. The corresponding
phases are Mott insulators. However for $\rho_{tot}=2$ only the correlated winding vanishes, meaning that there is no global flow of
particles, regardless of being atoms or molecules. But individual species are flowing, each in the opposite direction of the other, leading
to a large value of the anti-correlated winding. The phase is incompressible like a Mott insulator, but a supercurrent occurs for each of the species. We will refer to this phase
as "Super-Mott"\footnote{The name "Super-Mott" is chosen by analogy with the "supersolid" phase in the Bose-Hubbard model \cite{Fisher89}.}. We show in the following that this phase extends deep into the large-$g$ region of the phase diagram.

\subsection{Phase diagrams}
\begin{figure}[h]
  \centerline{\includegraphics[width=0.45\textwidth]{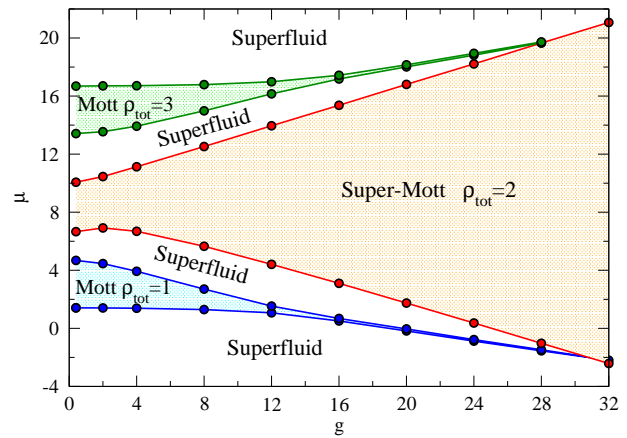}}
  \caption
    {
      (Color online) The phase diagram in the $(\mu,g)$ plane, for $U_{aa}=4$, $U_{am}=12$, and $D=6$.
    }
  \label{PhaseDiagramVsg}
\end{figure}

\begin{figure}[h]
  \centerline{\includegraphics[width=0.45\textwidth]{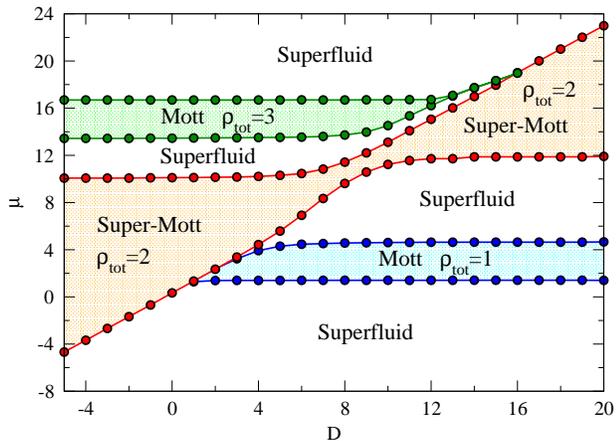}}
  \caption
    {
      (Color online) The phase diagram in the $(\mu,D)$ plane, for $U_{aa}=4$, $U_{am}=12$, and $g=0.5$.
    }
  \label{PhaseDiagramVsD}
\end{figure}

Finally, by reproducing Fig.\ref{RhoVsMu} for different sets of parameters $g$ and $D$ we are able to draw two phase diagrams,
one in the $(\mu,g)$-plane (Fig. \ref{PhaseDiagramVsg}) and one in the $(\mu,D)$-plane (Fig. \ref{PhaseDiagramVsD}). We can identify the three
incompressible phases discussed above, namely two Mott phases for $\rho_{tot}=1$ and $\rho_{tot}=3$, and the Super-Mott phase for $\rho_{tot}=2$.
Those phases extend over regions of the phase diagram separated by superfluid regions. For small $g$, all incompressible phases are present. As $g$ increases, the Super-Mott phase takes over the
two Mott phases (Fig. \ref{PhaseDiagramVsg}). For small or negative $D$ the Super-Mott phase takes over the $\rho_{tot}=1$ Mott phase, whereas
for large $D$ it is the $\rho_{tot}=3$ Mott phase which yields to the Super-Mott phase (Fig. \ref{PhaseDiagramVsD}).

\section{Summary and discussion}
We have studied a two-species Bose-Hubbard model including a conversion term between the two species. Our model can be of interest for ultracold atom experiments
using Feshbach resonances. The competition between the kinetic, potential, and conversion terms leads to rich phase diagrams. We have shown that increasing the
number of particles of the first species can lead to an inversion of population, resulting in the number of molecules greater than the number of atoms.
In addition to the usual superfluid and Mott phases occuring in boson models, we have identified an exotic "Super-Mott" phase, characterized by a vanishing
compressibility and a superflow of both species but with anticorrelations such that there is no global supercurrent. Finally, we have produced two phase diagrams
as a potential guide to detect the exotic Super-Mott phase. Since the Super-Mott phase occupies a big part of the phase diagrams, we expect it to be observable in
experiments. We are currently investigating the model using a newly developed algorithm \cite{SGF} that provides access to Green functions and momentum distribution functions,
which can be measured in experiments. This will allow a direct comparison between theory and experiments.

\begin{acknowledgments}
This work is part of the research program of the 'Stichting voor Fundamenteel Onderzoek der materie (FOM)', which is financially supported
by the 'Nederlandse Organisatie voor Wetenschappelijk Onderzoek (NWO)'. We would like to thank A.~Parson for his project.
\end{acknowledgments}

\end{document}